\documentclass{iopart}
\usepackage{graphicx}
\usepackage{latexsym}
\usepackage{amssymb}
\usepackage{verbatim}
\usepackage{iopams}

\begin{document}
\newcommand{\musr}{$\mu$SR{}}
\newcommand{\chem}[1]{\ensuremath{\mathrm{#1}}}

\title{Dynamic fields in the partial magnetization plateau of Ca$_3$Co$_2$O$_6$}

\author{P.\ J.\ Baker$^1$, J.\ S.\ Lord$^1$, and D.\ Prabhakaran$^2$}

\address{$^1$ ISIS Facility, STFC Rutherford Appleton Laboratory, Didcot, OX11 0QX, United Kingdom}

\address{$^2$ Oxford University Department of Physics, Clarendon Laboratory, 
Parks Road, Oxford OX1 3PU, United Kingdom}







\ead{peter.baker@stfc.ac.uk}

\date{\today}

\begin{abstract}
Fluctuation dynamics in magnetization plateaux are a relatively poorly explored area in frustrated magnetism. Here we use muon spin relaxation to determine the fluctuation timescale and associated field distribution width in the partial magnetization plateau of Ca$_3$Co$_2$O$_6$. The muon spin relaxation rate has a simple and characteristic field dependence which we model and by fitting to the data at $15$~K extract a fluctuation timescale $\tau = 880(30)$~ps and a field distribution width $\Delta = 40.6(3)$~mT. Comparison with previous results on Ca$_3$Co$_2$O$_6$ suggests that this fluctuation timescale can be associated with short-range, slowly fluctuating magnetic order.
\end{abstract}

\pacs{76.75.+i, 75.40.Gb, 75.50.Ee}
\submitto{JPCM}
\maketitle

\section{\label{sec:Introduction} Introduction}
When magnetic ions are arranged in certain structures their interactions can act to impede magnetic ordering. This is referred to as magnetic frustration and is widely studied~\cite{ramirez03,diep04}. A well known example of a frustrated structure is antiferromagnetically coupled spins on a triangular lattice~\cite{Collins97}. An alternative version of this is the antiferromagnetic coupling of ferromagnetic chains on a triangular lattice when the intrachain coupling is far stronger than the interchain coupling. This is realized in compounds such as CoNb$_2$O$_6$~\cite{Coldea10} and Ca$_3$Co$_2$O$_6$~\cite{Aasland97}, where the moments within the chains have Ising (easy-axis) anisotropy.  

In applied magnetic fields frustrated magnets can undergo transitions between different magnetic structures. In between such transitions the magnetization of the sample stays fixed over a significant range of fields: a magnetization plateau. These show a simple fraction of the saturation magnetization such as $1/3$ because the individual atomic spins are either aligned or anti-aligned with the external field. Such a partial magnetization plateau is observed in the intermediate, partially ordered phase of Ca$_3$Co$_2$O$_6$ between $10$ and $25$~K~\cite{Maignan04,Hardy04,Sugiyama05,Takeshita06,Fleck10}. A field of $\sim 0.5$~T is sufficient to enter the $1/3$ magnetization plateau and then a further transition into the saturated magnetic state occurs at $3.6$~T. Below $\sim 10$~K a series of magnetization steps are seen in single crystals for fields applied along the $c$-axis, with pronounced hysteresis~\cite{Hardy04,Fleck10}. That these magnetization steps occur in multiples of $1.2$~T has led to the suggestion that they represent quantum tunnelling of the magnetization~\cite{Maignan04,Hardy04}

Muon spin relaxation ($\mu$SR) can be used to investigate both the static and dynamic properties of frustrated magnets~\cite{blundell99,ddry,yddr}. This is most commonly carried out in zero applied field, since the technique is rare in allowing this and is exquisitely sensitive to small magnetic fields that could emerge due to weak magnetic order. It is sensitive to fluctuating magnetic fields on time scales ranging from around $10^{-11}$ to $10^{-5}$~s, intermediate between neutron scattering and ac magnetic susceptibility. Measurements on magnetic systems in applied fields along the initial direction of the muon spin allow the implanted muon spins to be decoupled from the field distribution they experience at their stopping site, providing a further window on the static and dynamic properties~\cite{ddry}. 

In this paper we describe the results of a $\mu$SR investigation of the magnetization plateau in Ca$_3$Co$_2$O$_6$. The broad and experimentally accessible plateau in the intermediate temperature region offers an ideal ground for quantitative comparison between models for the relaxation dynamics and experimental data. We propose a simple model to describe the field dependence of the muon spin relaxation rate, find that it describes the data successfully in the plateau region, and use it to estimate the time scale and field distribution associated with the magnetic fluctuations that the muon probes in this system.

\section{\label{sec:xto:exp} Experimental}

Stoichiometric Ca$_3$Co$_2$O$_6$ has been synthesised by the conventional solid state reaction technique.  High purity ($>99.99$~\%) starting materials CaCO$_3$ and Co$_3$O$_4$ were dried at $150^{\circ}$C for 6~h before weighing.  Mixed chemicals were calcined and sintered at $810-860^{\circ}$C for 48~h with intermediate grinding. Finally the sintered powder was pressed into cylindrical rods and sintered at $925^{\circ}$C for 48~h in air. 

\begin{figure}[htb]
\begin{center}
\includegraphics[width=\linewidth]{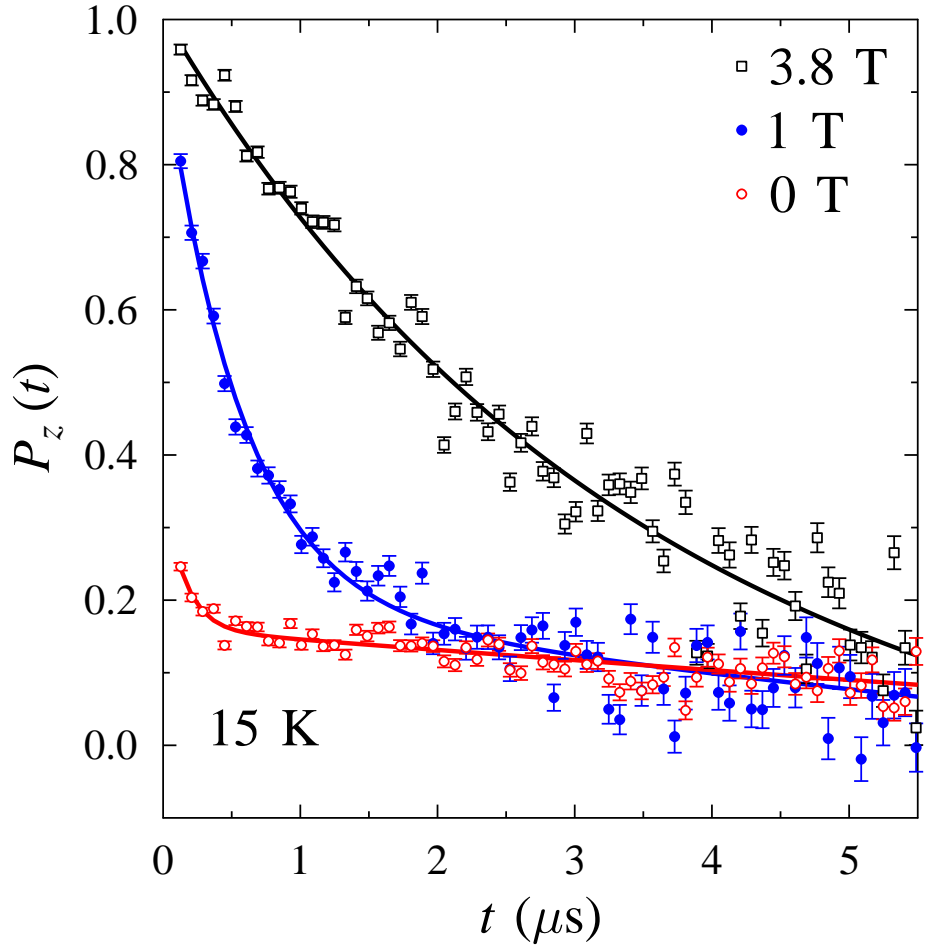}
\end{center}
\caption[Polarization spectra]{
Examples of the $P_{z}(t)$ data recorded at different fields for Ca$_3$Co$_2$O$_6$ at 15~K.
The lines plotted are fits of the data to Equation~{\ref{eq:fitfunc}}, adding a slowly relaxing background at 0~T.
}
\label{fig:data}
\end{figure}

Our $\mu$SR experiment~\cite{blundell99,ddry,yddr} was carried out on the newly constructed HiFi spectrometer~\cite{HiFi} at the ISIS Pulsed Muon Facility. This has a $5$~T superconducting magnet, almost ideally suited to the field range needed for this experiment. The powder sample was mounted in a 25~$\mu$m silver foil packet on a silver backing plate and cooled to $15$~K using a helium flow cryostat. Silver is used because the muon spin relaxation rate for muons stopping outside the sample will be close to zero over the field range of interest here.

The fully spin polarized muons are implanted into the sample and generally stop at interstitial positions within the crystal structure without significant loss of spin polarization. The spin polarization $P_{z}(t)$ of the muon depends on the magnetic environment of the of the muon stopping site and can be measured using the asymmetric decay of the muon into a positron, which is detected, and two neutrinos. Around 10 million decay positrons were detected in each data set. The emitted positrons are detected in scintillation counters around the sample position~\cite{HiFi}, divided into two banks forward (F) and backward (B) relative of the initial muon spin direction. The time-dependent asymmetry $A(t)$ between the count rates in the banks of detectors is:
\begin{equation}
A(t) = \frac{N_{\rm F}(t) - \alpha N_{\rm B}(t)}{N_{\rm F}(t) + \alpha N_{\rm B}(t)}, 
\label{eq:alpha}
\end{equation}
where the parameter $\alpha$ describes the relative counting efficiency of the two banks of detectors, depending on the sample and detector geometry. From $A(t)$ it is possible to infer the spin polarization $P_{z}(t)$ of the implanted muon ensemble:
\begin{equation}
P_{z}(t) = \frac{A(t) - A_{\rm bg}}{A(0)-A_{\rm bg}}. \label{eq:pz}
\end{equation}
The measured asymmetries $A(0)$ and $A_{\rm bg}$ are the initial and background values respectively and depend on the detector geometry and the fraction of the muon beam incident on the sample. The field variation of $\alpha$, $A(0)$, and $A_{\rm bg}$ required careful consideration in fields $\gtrsim 1$~T because the decay positrons spiral significantly
in the field. This was corrected for by comparison with data from reference samples~\cite{HiFi}. 

Following on from earlier muon work on Ca$_3$Co$_2$O$_6$ in the intermediate temperature region~\cite{Sugiyama05,Takeshita06}, we expect that in close to zero applied field the muon spin relaxation will not show the full initial asymmetry observed in the paramagnetic region because the internal fields are too large to be observed at ISIS. (The pulsed nature of the muon beam gives a range of arrival times for muons entering the sample which limits the maximum observed relaxation rate.) Takeshita and co-workers investigated the form of the spin relaxation up to around $1$~T in this temperature region, finding that the initial asymmetry had been recovered by $\sim 0.5$~T and that the relaxation was exponential over the whole field range. This agrees with the form of our data from $0$ up to $3.8$~T, as shown in Fig.~\ref{fig:data} and we therefore fitted our data to the function:
\begin{equation}
P_{z}(t) = \exp(-\lambda t),
\label{eq:fitfunc}
\end{equation}
where $\lambda$ is the relaxation rate. For $\lambda \gtrsim 0.1$~MHz, $\lambda$ can be extracted from the data independent from any field variation of $\alpha$, $A(0)$, and $A_{\rm bg}$. This is the case throughout our measured field range. In zero field $A_{\rm bg}$ decreases slowly with time due to muons stopping in the silver sample holder and any weak fields perpendicular to the initial muon spin direction, but in applied field $A_{\rm bg}$ is effectively constant.

To model the field dependence of $\lambda$ we make the assumption that the distribution of fluctuating magnetic fields approximates to a Lorentzian~\cite{ddry,yddr} with a width $\Delta$ and that one correlation time $\tau$ is an adequate description of the fluctuation rate. We further assume that within the partial magnetization plateau these two parameters can be taken to be constant. Making these assumptions allows us to simplify Redfield's equation~\cite{slichter} to the form:
\begin{equation}
\lambda = \frac{2\gamma^{2}_{\mu}\Delta^{2}\tau}{1+\gamma^{2}_{\mu}B^{2}_{\rm LF}\tau^2}.
\label{eq:redfield}
\end{equation}
The parameter $\gamma_{\mu} = 2\pi\times 135.5$~MHzT$^{-1}$ is the muon gyromagnetic ratio. 

\section{Results}
\label{sec:results}
\begin{figure}[htb]
\begin{center}
\includegraphics[width=0.5\linewidth]{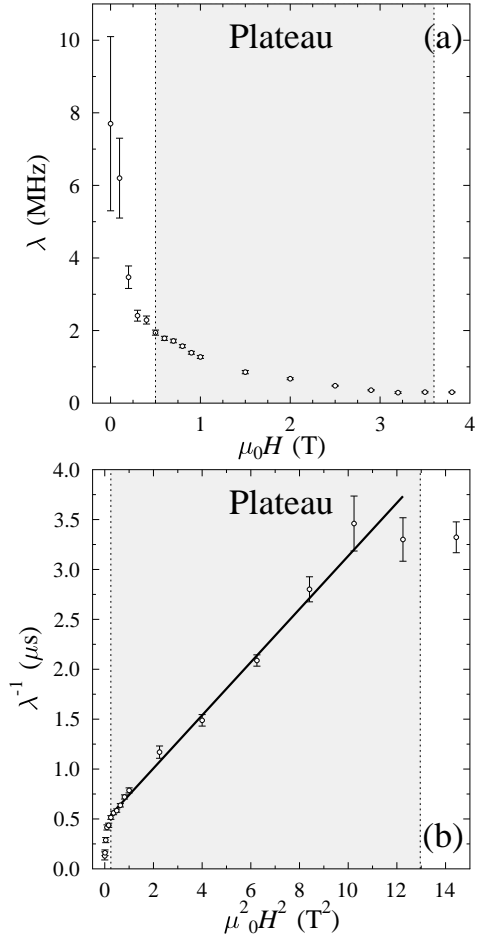}
\end{center}
\caption[Parameters extracted from \musr{} data]{
Parameters extracted from the \musr{} data using Equation~{\ref{eq:fitfunc}}: 
(a) Relaxation rate $\lambda$ plotted against the applied field $\mu_0 H$. 
(b) Inverse of the relaxation rate $\lambda^{-1}$ plotted against the square of the applied field $\mu^{2}_0 H^2$. The trend obtained by fitting the data to Equation~\ref{eq:redfield} is shown in the shaded magnetization plateau region.
}
\label{fig:results}
\end{figure}

In Fig.~\ref{fig:results}~(a) we show the relaxation rate $\lambda$ extracted from the data at 15~K as a function of the magnetic field $B_{\rm LF}=\mu_0 H$. Three behaviours are apparent in the data: At fields below $0.5$~T there is a rapid decrease in $\lambda$, between $0.5$ and $3.6$~T the decrease in $\lambda$ with increasing field is slower, and above $3.6$~T any change in $\lambda$ is smaller than the experimental error. These regions correspond to the initial increase in bulk magnetization, the magnetization plateau, and the saturated magnetic phase respectively.

To make a more obvious comparison with the Redfield equation~(\ref{eq:redfield}) discussed above we plot $\lambda^{-1}$ against $\mu^{2}_0 H^2$ in Fig.~\ref{fig:results}~(b). This rearrangement means that the trend for constant $\Delta$ and $\tau$ will appear as a straight line, as is observed in the data in the magnetization plateau region highlighted.
Fitting equation~\ref{eq:redfield} to the data shown in Fig.~\ref{fig:results} between $0.5$ and $3.6$~T gives the parameters: $\Delta = 40.6(3)$~mT and $\tau = 880(30)$~ps. This gives the line plotted in Fig.~\ref{fig:results}~(b). These parameters are insensitive to small reductions in the fitting range at either end of the plateau region.

\section{Conclusion}
\label{sec:conclusion}
Using the partial magnetization plateau in Ca$_3$Co$_2$O$_6$ as a model system in which to test a simple model for the field-dependence of the muon spin relaxation rate in a magnetization plateau we find very good agreement between the form predicted by the model and the experimental data. We are able to obtain a field distribution width and fluctuation timescale for the dynamic magnetism within the plateau. This suggests that this model could also be applied to other magnetic systems that exhibit a magnetization plateau and provide new information about their dynamic behaviour.

The width of the field distribution that we obtain is comparable with the $\sim 100$~mT static internal field observed in zero-field measurements~\cite{Sugiyama05}. However, were the muon probing such static fields they would have been decoupled from the muon spin at fields lower than those in the magnetization plateau. Both the previous $\mu$SR results~\cite{Sugiyama05} and neutron diffraction work~\cite{Fleck10} show strong evidence for shorter ranged, dynamic correlations coexisting with the long-range magnetic order at this temperature. Such dynamic correlations would not be decoupled from the muon spin so easily and it is therefore this behaviour that we are probing in this experiment. Since the field distribution width will be determined by the same magnetic moments the same distance from the muon stopping site(s), and this is entirely consistent with our results.

The timescale we find for the fluctuations $\lesssim 1$~ns sits comfortably within the sensitivity range of $\mu$SR and suggests that these fluctuations would be static on the timescale of neutron diffraction. Again, this is consistent with the previous neutron diffraction measurements where the short-range magnetic order is quasistatic~\cite{Fleck10}, and the previous $\mu$SR results where a significant strongly-relaxing component suggestive of slowing fluctuating short-ranged order is observed at this temperature~\cite{Sugiyama05,Takeshita06}. Earlier ac susceptibility measurements found a significantly longer timescale, $\sim 0.4$~s~\cite{Hardy04}, which is far outside the window we can probe using muons, and is likely to be associated with a separate process within the system.

Comparing our model to the data in the region $B < 0.5$~T we find far poorer agreement. Of course, the assumption that $\Delta$ is independent of field is unlikely to be appropriate here and we would only get some approximation to the average value. The parameters coming from our data for $B < 0.5$~T are: $\Delta \simeq 30$~mT and $\tau \simeq 5$~ns, which suggest that the decoupling from longer ranged order remains relevant in this field range. Takeshita {\em et al.} carried out a similar analysis of their data at 20~K and obtained larger, although comparable, values for both parameters~\cite{Takeshita06}.

\section{Acknowledgements}
\label{sec:acknowledgements}
We thank the Science and Technology Facilities Council for a Facility Development grant (FDPG/082) and the European Commission under the 7$^{\rm th}$ Framework Programme through the Key Action: Strengthening the European Research Area, Research Infrastructures: Contract no: CP-CSA INFRA-2008-1.1.1 Number 226507-NMI3. We acknowledge EPSRC support for sample growth. Data were collected with assistance from: Roxana Dudric, Hassan Saadaoui, Robert Tarasenkko, Lee Iverson, Calin Rusu, Adam Berlie, Edwin Kermarrec, and Neda Nikseresht.

\end{document}